\documentclass[12pt]{article}
\usepackage{graphicx}
\usepackage{amsmath}
\usepackage{amssymb}
\usepackage{hyperref}

\newcommand\pubdate{\today}

\def\Title#1{\begin{center} {\Large #1 } \end{center}}
\def\Author#1{\begin{center}{ \sc #1} \end{center}}
\def\Address#1{\begin{center}{ \it #1} \end{center}}

\newcommand\pubblock{\rightline{\begin{tabular}{l}  \\ 
         \pubdate  \end{tabular}}}
\newenvironment{Abstract}{\begin{quotation}  }{\end{quotation}}
\newenvironment{Presented}{\begin{quotation} \begin{center}
             PRESENTED AT\end{center}\bigskip
      \begin{center}\begin{large}}{\end{large}\end{center} \end{quotation}}

\begin{document}
\begin{titlepage}
 \pubblock
\vfill
\Title{Quarkonia pair production as a tool for study of gluon GPDs}
\vfill
\Author{Marat Siddikov, Ivan Schmidt}
\vfill

\Address{Departamento de Física, Universidad Técnica Federico Santa María, y Centro Científico - Tecnológico de Valparaíso, Casilla 110-V, Valparaíso, Chile}
\vfill
\begin{Abstract}
In these proceedings we present our results on the exclusive photoproduction of $J/\psi\,\eta_{c}$ pairs in the collinear factorization framework. We argue that the process might be used as a complementary channel for studying the generalized parton distributions (GPDs) of gluons. We provide numerical estimates for the cross-section in the kinematics of the future Electron Ion Collider.
\end{Abstract}
\vfill
\begin{Presented}
DIS2023: XXX International Workshop on Deep-Inelastic Scattering and
Related Subjects, \\
Michigan State University, USA, 27-31 March 2023 \\
     \includegraphics[width=9cm]{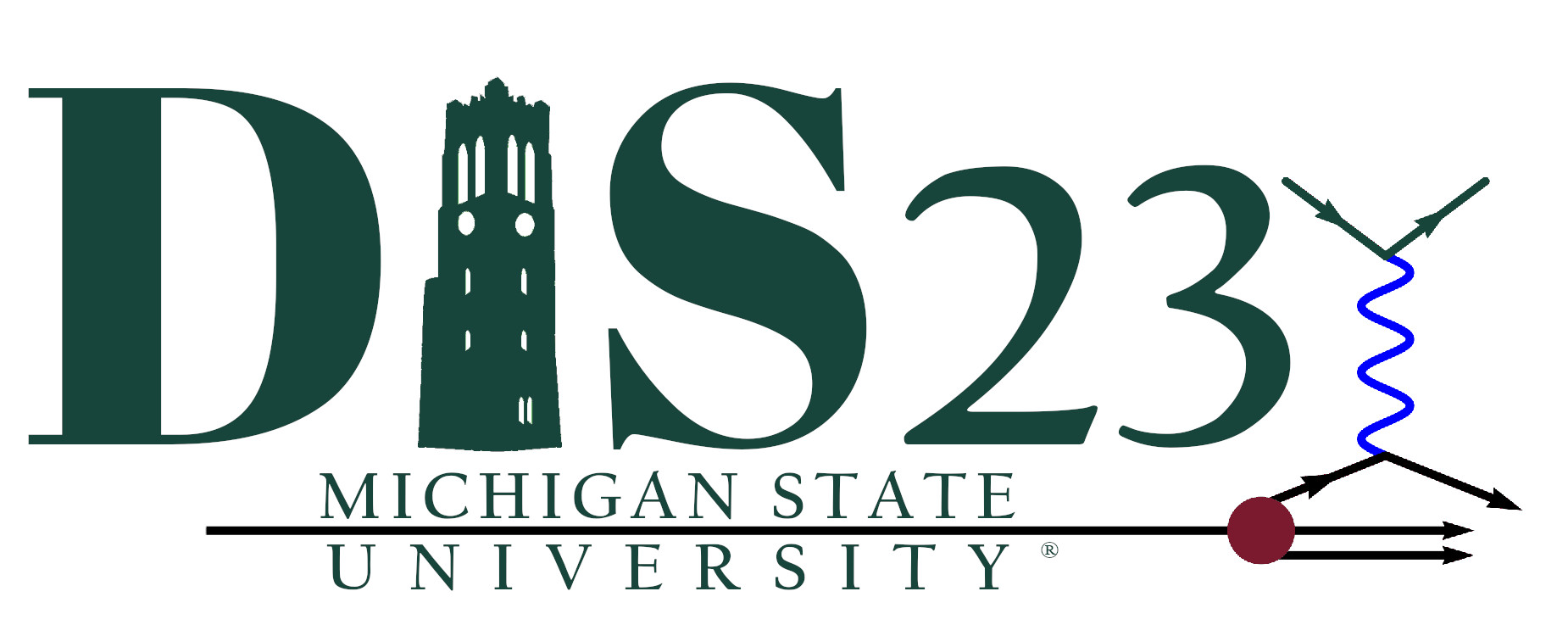}
\end{Presented}
\vfill
\end{titlepage}

\section{Introduction}

Nowadays the understanding of partonic and multipartonic distributions in the proton, and in particular of the so-called Generalized Parton Distributions (GPDs)~\cite{Goeke:2001tz,Diehl:2003ny,Guidal:2013rya}
remains one of the  open problems within hadronic physics. The phenomenological extraction of these distributions is challenging for technical (mathematical) reasons. Moreover, it relies on different assumptions and sometimes provides only limited information about
the partonic distributions. For these reasons, it is always desired
to extend the number of channels used for phenomenological studies~\cite{Pire:2017yge,Pire:2021dad}. Recently it has been suggested that $2\to3$ exclusive processes might be used as a new tool for the study of  the GPDs and complement existing phenomenological
research in $2\to2$ channels~\cite{GPD2x3:9,GPD2x3:7,GPD2x3:6,GPD2x3:5,GPD2x3:4,Boussarie:2016qop,GPD2x3:10}.
We argue that it is possible to extend these studies and use the exclusive photoproduction of heavy quarkonia pairs as additional probe of the gluon GPDs. From previous research on single-quarkonia production it is known that the heavy mass of quarkonia may be used as a natural hard scale in the problem, justifying the use of perturbative methods, even in the photoproduction regime.
Due to the different structure of the coefficient function, the quarkonia
\emph{pair} production allows to get additional constraints on the gluon GPDs,
especially outside the classical $x=\pm\xi$ line.
Since,  due to $C$-parity constraints, the production of $J/\psi\,J/\psi$ pairs
is not related to GPDs of the target, in our
study we focus on the production of $J/\psi\,\eta_{c}$ pairs, and
analyze the kinematics of the low-energy runs at the future Electron-Ion
Collider~\cite{Accardi:2012qut,AbdulKhalek:2021gbh}.
For high-energy runs, as well as other future accelerators, it might
be more appropriate to use the evaluations in the color dipole picture,
which incorporates saturation effects~\cite{Andrade:2022rbn}.

This proceeding is structured as follows. Below, in Section~\ref{sec:Formalism},
we briefly discuss  the  framework and the structure of the cross-section
of quarkonia pair production (the detailed derivation of these results
might be found in~\cite{Siddikov:2022bku}). At the end of this section
we present some numerical estimates for the cross-section in the EIC kinematics and
 draw conclusions.

\section{Exclusive photoproduction of meson pairs}

\label{sec:Formalism}

The production of \emph{light} meson pairs was analyzed previously
in Refs.~\cite{LehmannDronke:2000hlo,Clerbaux:2000hb},
in Bjorken kinematics. However, for heavy quarkonia this analysis has limited applicability, since in the kinematics of very large photon virtualities
$Q^{2}=-q^{2}\gg M_{1,2}^2$ (quarkonia masses), the cross-section is vanishingly small. In our studies we consider that both $Q^{2}$ and $M_{1,2}^{2}$ are large scales, although eventually we will consider the photoproduction
limit $Q\to0$.

The cross-section of the photoproduction of heavy quarkonia pairs is given by
\begin{equation}
d\sigma_{\gamma p\to M_{1}M_{2}p}^{(L,T)}=\frac{dy_{1}dp_{1\perp}^{2}dy_{2}dp_{2\perp}^{2}d\phi\left|\mathcal{A}_{\gamma p\to M_{1}M_{2}p}^{(L,T)}\right|^{2}}{4\left(2\pi\right)^{4}\sqrt{\left(W^{2}+Q^{2}-m_{N}^{2}\right)^{2}+4Q^{2}m_{N}^{2}}}\delta\left(\left(q+P_{1}-p_{1}-p_{2}\right)^{2}-m_{N}^{2}\right)\label{eq:Photo-2}
\end{equation}
where $y_{1},y_{2}$ are the quarkonia rapidities, $p_{1\perp},\,p_{2\perp}$
are their corresponding momenta, $\phi$ is the azimuthal angle between
$\boldsymbol{p}_{1\perp},\boldsymbol{p}_{2\perp}$; $W^{2}=\left(q+P_1\right)^{2}$
is the energy of the $\gamma^{*}p$ pairs, and the $\delta$-function
in the right-hand side stems from the onshellness of the recoil proton.
This $\delta$-function introduces cumbersome constraints on the kinematics
of the produced quarkonia pairs for the \emph{fixed-energy} photons
(see~\cite{Siddikov:2022bku} for details), however might be trivially
taken into account if we treat the quarkonia momenta $p_{1\perp},\,p_{2\perp}$
and rapidities $y_{1},y_{2}$ as independent variables, and fix the
photon energy $W$ from the onshellness condition.

\begin{figure}
\includegraphics[height=6cm]{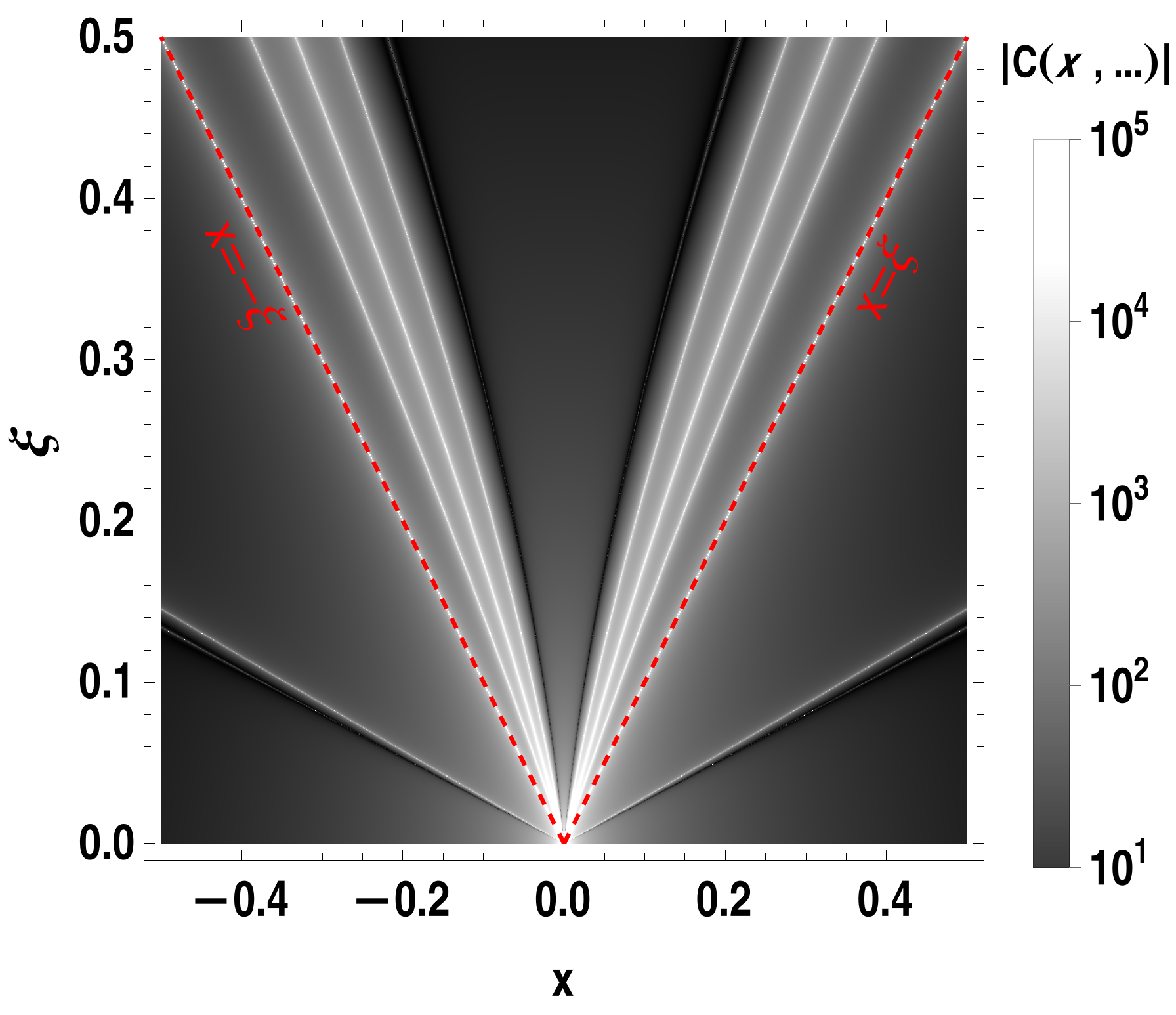}{\footnotesize{}\includegraphics[height=6cm]{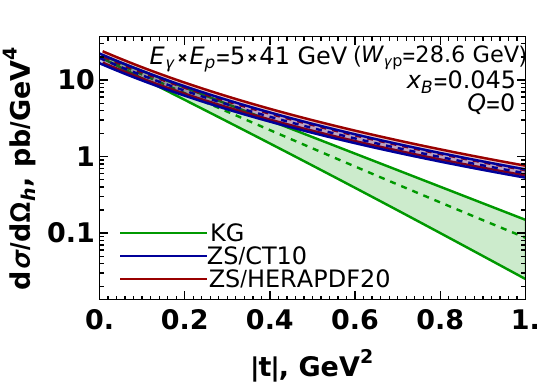}}
\caption{\label{fig:CoefFunction} Left: The coefficient function $c_{T}\left(x,\,y_{1}=y_{2}\right)$
for transversely polarized photons, as a function of the variables
$x$ and skewness variable $\xi$, for the case of photoproduction
($Q=0$) and equal quarkonia rapidities $y_{1}=y_{2}$. The bright lines effectively demonstrate the position of the poles $x_{k}^{\ell}$ of the coefficient function~(\ref{eq:Monome-1}).
For reference, we also marked with red dashed lines the position of
 \textquotedblleft classical\textquotedblright{} poles $x=\pm\xi$,
which are probed in DVCS and DVMP. Right: Dependence of the cross-section
of the photoproduction of $J/\psi\,\eta_{c}$ on the momentum transfer
$t$ to the proton for the low-energy EIC beam, assuming equal rapidities
$y_{1}=y_{2}$ of the two quarkonia. Various colored bands differ
only due to choice of the gluon GPDs: KG stands for Kroll-Goloskokov
parametrization~\cite{Goloskokov:2013mba}, ZS is the so-called zero skewness parametrization  $H_{ZS}^{g}\left(x,\xi,t, \mu^2 \right)=g\left(x,\mu^{2}\right)F(t)$, supplemented with different choices of gluon PDFs $g\left(x,\mu^{2}\right)$. The width of the colored band reflects
the uncertainty due to higher order corrections, which is estimated
varying the factorization scale $\mu_{F}$ between $\mu_{F}=2M_{J/\psi}$
and $\mu_{F}=M_{J/\psi}/2$ respectively (the central dashed line
corresponds to $\mu=M_{J/\psi}$). }
\end{figure}

The evaluation of the amplitudes $\mathcal{A}_{\gamma p\to M_{1}M_{2}p}^{(L,T)}$
was done in the collinear factorization framework, assuming the quarkonia
pairs and the recoil proton are kinematically well-separated from
each other. In the leading order, the dominant contribution to the
amplitudes of quarkonia production comes from the gluon GPDs. In our
evaluations we will disregard the contributions of the poorly known
transversity gluon GPDs $H_{T}^{g},\,E_{T}^{g},\,\tilde{H}_{T}^{g},\,\tilde{E}_{T}^{g}$,
since existing experimental bounds suggest that they should be negligibly
small (see e.g. explanation in~\cite{Pire:2017yge,Goloskokov:2013mba}).
The contribution of the remaining (chiral even) GPDs to the square
of amplitude is given by

\begin{align}
&\sum_{{\rm spins}}\left|\mathcal{A}_{\gamma p\to M_{1}M_{2}p}^{(\mathfrak{a})}\right|^{2} =\frac{1}{\left(2-x_{B}\right)^{2}}\left[4\left(1-x_{B}\right)\left(\mathcal{H}_{\mathfrak{a}}\mathcal{H}_{\mathfrak{a}}^{*}+\tilde{\mathcal{H}}_{\mathfrak{a}}\tilde{\mathcal{H}}_{\mathfrak{a}}^{*}\right)\right.\label{eq:AmpSq-1}\\
&\qquad-x_{B}^{2}\left(\mathcal{H}_{\mathfrak{a}}\mathcal{E}_{\mathfrak{a}}^{*}+\mathcal{E}_{\mathfrak{a}}\mathcal{H}_{\mathfrak{a}}^{*}+\tilde{\mathcal{H}}_{\mathfrak{a}}\tilde{\mathcal{E}}_{\mathfrak{a}}^{*}+\tilde{\mathcal{E}}_{\mathfrak{a}}\tilde{\mathcal{H}}_{\mathfrak{a}}^{*}\right)\nonumber\\
 & \qquad\left.-\left(x_{B}^{2}+\left(2-x_{B}\right)^{2}\frac{t}{4m_{N}^{2}}\right)\mathcal{E}_{\mathfrak{a}}\mathcal{E}_{\mathfrak{a}}^{*}-x_{B}^{2}\frac{t}{4m_{N}^{2}}\tilde{\mathcal{E}}_{\mathfrak{a}}\tilde{\mathcal{E}}_{\mathfrak{a}}^{*}\right],\qquad\mathfrak{a}=L,T\nonumber
\end{align}
where we introduced the shorthand notations for convolutions

\begin{align}
\mathcal{H}_{\mathfrak{a}} & =\int_{-1}^{1}dx\,c_{\mathfrak{a}}\left(x,\,y_{1},\,y_{2}\right)H_{g}\left(x,\xi,t\right),\quad\mathcal{E}_{\mathfrak{a}}=\int_{-1}^{1}dx\,c_{\mathfrak{a}}\left(x,\,y_{1},\,y_{2}\right)E_{g}\left(x,\xi,t\right),\label{eq:Ha-1}\\
\tilde{\mathcal{H}}_{\mathfrak{a}}& =\int_{-1}^{1}dx\,\tilde{c}_{\mathfrak{a}}\left(x,\,y_{1},\,y_{2}\right)\tilde{H}_{g}\left(x,\xi,t\right),\quad\tilde{\mathcal{E}}_{\mathfrak{a}}=\int_{-1}^{1}dx\,\tilde{c}_{\mathfrak{a}}\left(x,\,y_{1},\,y_{2}\right)\tilde{E}_{g}\left(x,\xi,t\right),\label{eq:ETildeA-1}
\end{align}
$x$ is the average light-cone momentum fraction of the proton carried
by the gluon before and after interaction, and $\xi$ is the standard
skewness variable (it might be related to quarkonia rapidities $y_{1},y_{2}$).
The partonic amplitudes $c_{\mathfrak{a}},\,\tilde{c}_{\mathfrak{a}}$
might be evaluated perturbatively (see details in~\cite{Siddikov:2022bku}). For the case in which the quarkonia are well-separated from each other kinematically,
it is possible to express the amplitudes $c_{\mathfrak{a}},\,\tilde{c}_{\mathfrak{a}}$
in terms of the  nonperturbative long-distance matrix elements (LDMEs)
of Non-Relativistic QCD (NRQCD)~\cite{Brambilla:2010cs}, multiplied by a rational function,
\begin{equation}
c_{\mathfrak{a}},\,\tilde{c}_{\mathfrak{a}}\sim\sum_{\ell}\frac{\mathcal{P}_{\ell}\left(x\right)}{\prod_{k=1}^{n_{\ell}}\left(x-x_{k}^{(\ell)}+i0\right)}\label{eq:Monome-1}
\end{equation}
where $\mathcal{P}_{\ell}\left(x\right)$ is a smooth polynomial of
the variable $x$, and the denominator of each term in the sum~(\ref{eq:Monome-1})
might include a polynomial with up to $n_{\ell}=5$ nodes $x_{k}^{(\ell)}$
in the region of integration. The position of the poles $x_{k}^{(\ell)}$
depends on all kinematic variables $y_{1},\,y_{2},\,Q$, and for this reason, varying the rapidities $y_{1},y_{2}$ of the observed quarkonia, it is possible to probe the gluon GPDs in the full kinematic range $(x,\,\xi)$.
Due to space limitations, here we omit the full expressions for the
amplitudes $c_{\mathfrak{a}},\,\tilde{c}_{\mathfrak{a}}$ (see~\cite{Siddikov:2022bku} for details). However, in 
Figure~\ref{fig:CoefFunction}
we show the density plot which illustrates the behavior of the coefficient
function $c_{T}\left(x,\,y_{1},\,y_{2}\right)$ as a function of its
arguments, and allows to see the dependence of the poles on the variable $\xi$.

The typical values of the cross-sections in the EIC kinematics range
between a few dozens to a few hundreds of picobarns, depending on
the kinematics and chosen parametrization of the gluon GPDs.
In the right panel of Figure~\ref{fig:CoefFunction} for the
sake of illustration we show the cross-section for the lowest-energy
electron-proton beam as a function of the invariant momentum transfer
$t$, for several parametrizations of the gluon GPDs.
More detailed predictions for the cross-section at various
energies might be found in~our recent article~\cite{Siddikov:2022bku}.

To summarize, our findings demonstrate that the\emph{ }exclusive photoproduction
of $J/\psi\,\eta_{c}$ mesons (as well as other heavy quarkonia pairs
with opposite $C$-parities) potentially could be used as a viable
gateway for the analysis of the gluon GPDs of the target. The amplitude
of this process obtains the dominant contribution from the unpolarized
gluon GPD $H_{g}$; however, in contrast to classical $2\to2$ processes,
it has sensitivity to the behavior of the GPDs outside the $x=\pm\xi$
line, and thus could complement information extracted from DVCS and
single-quarkonia production. Numerically, the evaluated cross-sections are on par with similar
estimates for $2\to3$ processes suggested recently in the literature~\cite{GPD2x3:9,GPD2x3:7,GPD2x3:6,GPD2x3:5,GPD2x3:4}.

\section*{Acknowledgements}

We thank our colleagues at UTFSM university for encouraging discussions.
This research was partially supported by Proyecto ANID PIA/APOYO AFB220004
(Chile) and Fondecyt (Chile) grants 1220242 and 1230391. ``Powered@NLHPC:
This research was partially supported by the supercomputing infrastructure
of the NLHPC (ECM-02)''.

 \end{document}